# Dualheap Selection Algorithm:
# Efficient, Inherently Parallel and Somewhat Mysterious

Greg Sepesi
Eduneer, LLC
sepesi@eduneer.com
2007 June 14

**ABSTRACT**

An inherently parallel algorithm is proposed that efficiently performs selection: finding the K-th largest member of a set of N members. Selection is a common component of many more complex algorithms and therefore is a widely studied problem.

Not much is new in the proposed dualheap selection algorithm: the heap data structure is from J.W.J. Williams [1], the bottom-up heap construction is from R.W. Floyd [2], and the concept of a two heap data structure is from J.W.J. Williams and D.E. Knuth [3]. The algorithm's novelty is limited to relatively minor implementation twists:

- the two heaps are oriented with their roots at the partition values rather than at the minimum and maximum values,
- the coding of one of the heaps (the heap of smaller values) employs negative indexing,
- the exchange phase of the algorithm is similar to a bottom-up heap construction, but navigates the heap with a post-order tree traversal.

When run on a single processor, the dualheap selection algorithm's performance is competitive with quickselect with median estimation, a common variant of C.A.R. Hoare's quicksort algorithm [4]. When run on parallel processors, the dualheap selection algorithm is superior due to its subtasks that are easily partitioned and innately balanced.

## 1. ALGORITHM OVERVIEW

A heap is an array with elements regarded as nodes in a complete binary tree, where node j is the parent of nodes 2j and 2j+1, and where the value at each parent node is superior to the values at its children's nodes. This superiority of all the parent nodes is commonly called the heap condition.

The dualheap selection algorithm consists of three phases that are roughly equivalent in terms of the number of comparisons and moves they perform:

1) the whole heap construction phase,
2) the split heap construction phase,
3) the exchange phase.

**Phase 1** is a bottom-up heap construction as described in many algorithm textbooks such as "Algorithms" by Sedgewick [5]. Although not strictly required, this initial bottom-up heap construction typically reduces the total number of comparison and move operations by about 10%.

**Phase 2** consists of two more bottom-up heap constructions that split the original set of N members into a size K heap and a size N - K heap. It is worth noting that the partition point in the dualheap selection algorithm is set just once, based upon the selected values of N and K rather than the values being partitioned.

**Phase 3** repeatedly exchanges subtrees between the two heaps until all the values in the size K heap of larger values are not smaller than any of the values in the size N - K heap of smaller values.

Symbolically, the downward pointing triangle in Figure 1-1 represents the size K heap of larger values, the upward pointing triangle represents the size N - K heap of smaller values, and the arrow shows the direction of increasing values and increasing heap indices, indicating that the size N - K heap of smaller values employs negative indices. Phase 3 keeps exchanging values between the two heaps until the heap value ranges no longer overlap. At that point, the K-th largest member of the original set of N is at the root of the size K heap of larger values.

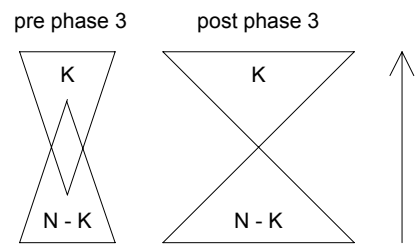

**Figure 1-1. Dualheap (Pre and Post Phase 3)**

## 2. ALGORITHM DETAILS

The dualheap selection algorithm is based upon downheap, originally proposed by J.W.J. Williams [1], that "promotes" the root of a subheap as far as possible down the subheap's greedy path. For example, suppose the size K heap of larger values, shown symbolically in Figure 1-1 as the downward pointing triangle, has a value (temporarily) at its root that is larger than the value of at least one of its children. Clearly, this root of the size K heap of larger values is violating the heap condition and needs to be exchanged with some smaller value. The downheap task elegantly accomplishes the exchange by shifting (i.e., promoting) the root value as far as possible down the greedy path of small children until the value finally reaches a node at which it is smaller than both the children there. The greedy path of small children starts at the root and follows the path down the heap by navigating at each level to the smaller of the two children. The significance of being a greedy path is that it allows the portion of the path down to the new position of the promoted value to shift up into the vacated root position without changing the heap condition. (Each new parent in that portion of the greedy path was previously the smaller sibling and is therefore now smaller than both its current children.)

As originally proposed by R.W. Floyd [2] and shown in Listing 2-1, the bottom-up construction of a heap executes the downheap task at each node encountered while decrementing through the nodes of the heap (starting at node N / 2 because nodes beyond node N / 2 have no children). In the listing, the 'S' and 'L' postfixes denote the small valued heap and the large valued heap, respectively. For example, NS is the size of the small valued heap (i.e., N – K) and NL is the size of the large valued heap (i.e., K). The process is called bottom-up heap construction because it constructs the smallest subheaps first then builds successively larger subheaps on top of them.

```
Phase1() {
   int i;
   for (i=N/2; i>0; i--)
      DownHeap(i);
}

Phase2() {
   int i;
   for (i=NS/2; i>0; i--)
      DownHeapS(i);
   for (i=NL/2; i>0; i--)
      DownHeapL(i);
}
```

**Listing 2-1. Phases 1 and 2 – Bottom-Up Heap Construction**

The worst case performance of bottom-up heap construction is linear time. That worst case occurs when the input is such that each of the N / 2 nodes traversed during the bottom-up heap construction has a value that needs to be promoted all the way to the bottom of its subheap. The maximum promotion distance for any subheap is the subheap's height, which for a size N heap is int(lg(N)). The total maximum promotion distance for all the subheaps is N – int(lg(N)) – 1, which of course is linear time [6]. (Note that the number of comparisons is twice the number of promotions because each promotion step requires one comparison to determine the next step of the greedy path and another comparison to determine if the promotion should take that step.) Therefore, the worst case performances of the first two phases of the dualheap selection algorithm are linear time, as they are simply bottom-up heap constructions.

Phase 3 is similar to a bottom-up heap construction in that they both perform the downheap task on nodes visited during an orderly traversal of the heap. For the bottom-up heap construction, the orderly traversal is simply decrementing through the nodes. And for phase 3, it is a post-order tree traversal. Typically post-order traversal is described as "visit the left subtree, visit the right subtree, then visit the root." However for phase 3, it is "visit the greedy subtree, visit the not greedy subtree, then visit the root." It is worth noting that this depth first traversal is guaranteed to use no more than 2 lg(N) stack frames..

Listing 2-2 shows how phase 3 repeatedly calls TreeSwap until the root of the size K heap of larger values is not smaller than the root of the size N – K heap of smaller values (i.e., until the heap value ranges no longer overlap). As with Listing 2-1, the 'S' and 'L' postfixes denote the small valued heap and the large valued heap, respectively. For example, pS is a pointer to the size N – K heap of small values, jS is an index into it, and DownHeapS operates on it. It is worth noting that the pointers pS and pL point to element zero of the heaps and the roots of the heaps are at element one (i.e., pS[-1] and pL[1]).

```
TreeSwap(int kS, int kL) {
   int jS, jL, tmp;

   jS = 2*kS;
   jL = 2*kL;
   if ((jS<=NS) && (jL<=NL)) {
      jS += (pS[-jS-1] > pS[-jS])?1:0;
      jL += (pL[jL+1] < pL[jL])?1:0;
      if (pS[-jS] > pL[jL]) {
         TreeSwap(jS, jL);
         if (pS[-(jS^1)] > pL[jL^1])
            TreeSwap(jS^1, jL^1);
      }
   }
   tmp = pS[-kS];
   pS[-kS] = pL[kL];
   pL[kL] = tmp;
   DownHeapS(kS);  DownHeapL(kL);
}

Phase3() {
   while (pS[-1] > pL[1])
      TreeSwap(1,1);
}
```

**Listing 2-2.  Phase 3 – Tree Swap**

Listing 2-3 shows the versions of downheap: DownHeapL operates on the size K heap of larger values and DownHeapS operates on the size N – K heap of smaller values.

```
void DownHeapS(int k) {
   int j, v;

   if (k <= NS/2) {
      v = pS[-k];
      for (j=2*k,
         j+=(pS[-j-1] > pS[-j])?1:0;
         pS[-j] > v;
         k=j, j=2*k,
         j+=(pS[-j-1] > pS[-j])?1:0)
      {
         pS[-k] = pS[-j];
         if (j > NS/2) {
            k = j;
            break;
         }
      }
      pS[-k] = v;
   }
}
```

```
void DownHeapL(int k) {
   int j, v;

   if (k <= NL/2) {
      v = pL[k];
      for (j=2*k,
         j+=(pL[j+1] < pL[j])?1:0;
         pL[j] < v;
         k=j, j=2*k,
         j+=(pL[j+1] < pL[j])?1:0)
      {
         pL[k] = pL[j];
         if (j > NL/2) {
            k = j;
            break;
         }
      }
      pL[k] = v;
   }
}
```

**Listing 2-3.  DownHeap**

Although TreeSwap and bottom-up heap construction are similar in that they both traverse the heap in a bottom-up manner while calling DownHeap(), TreeSwap is more mysterious:

- **What governs the convergence?**  Phase 3 always converges.  Each execution of TreeSwap is a figurative turn of the crank that adds more constraints to the heaps and eventually forces their separation.  However little is known about the process (e.g., how often do values get multiple transfers between the heaps?).  Figure 2-4 illustrates how the number of calls to TreeSwap varies with N over 32,000 test cases that equally partition uniformly distributed and pseudo-randomly generated input.  It shows, for instance, that typically fewer than 10 TreeSwaps are required to partition arrays of 10,000 elements in half.  What are the bounds to the number of calls to TreeSwap?

- **What is the worst case performance?** Unlike bottom-up heap construction that operates mostly on small subheaps, TreeSwap operates mostly on large subheaps. And unlike bottom-up heap construction that needs only one pass, TreeSwap typically needs many passes.  Given these differences, one might not expect for bottom-up heap construction and phase 3 to both have worst case linear-time performance.  But apparently they do.

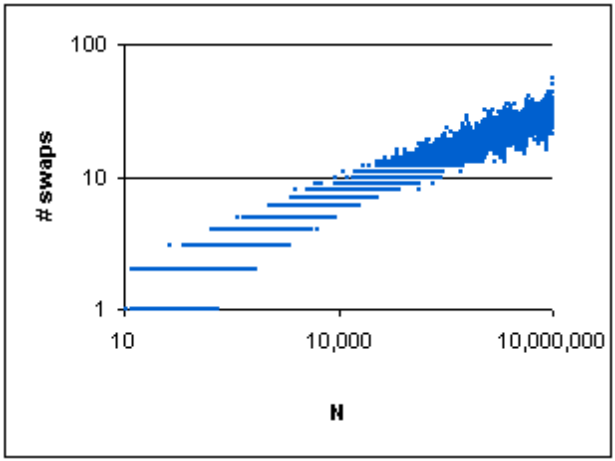

Figure 2-4. Number of Tree Swaps

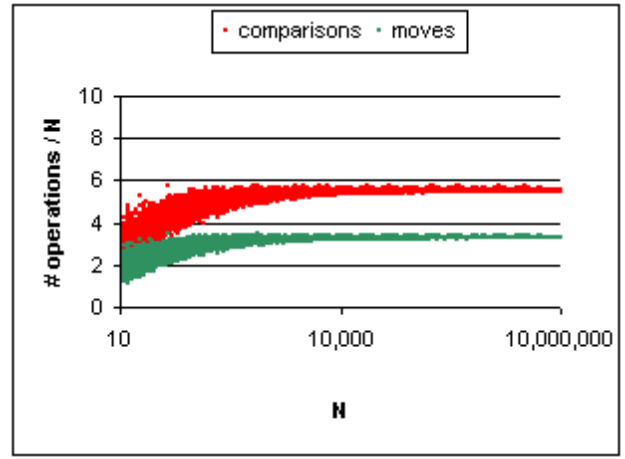

Figure 3-1. Operations by Dualheap Selection

## 3. EMPIRICAL COMPARISON

Figures 3-1, 3-2 and 3-3 illustrate the typical performance of the dualheap selection, quickselect with median estimate and quickselect algorithms respectively, by showing the number of comparison and move operations performed by 32,000 test cases that equally partition uniformly distributed and pseudo-randomly generated input.

Comparing Figures 3-1 and 3-3, quickselect typically performs fewer comparison and move operations. Its inner loop is also simpler and faster than the dualheap selection algorithm's inner loop (i.e., DownHeap). Although quickselect typically outperforms the dualheap selection algorithm, quickselect has a notorious worst case performance (i.e., $O(N^2)$). [7]

Robust implementations of quickselect employ some enhancement that avoids using consistently poor (i.e., unbalanced) partition values that lead to quickselect's worst case performance. One common enhancement is to set the partition value to a reliable estimate of the median. Unfortunately, adding any of these enhancements slows down the performance of quickselect. Comparing Figures 3-1 and 3-2, dualheap selection typically performs significantly fewer comparison and move operations than a robust implementation of quickselect.

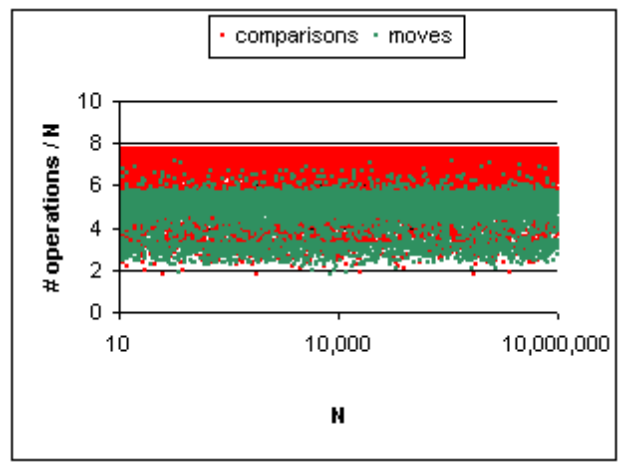

Figure 3-2. Operations by Quickselect with Median Estimate

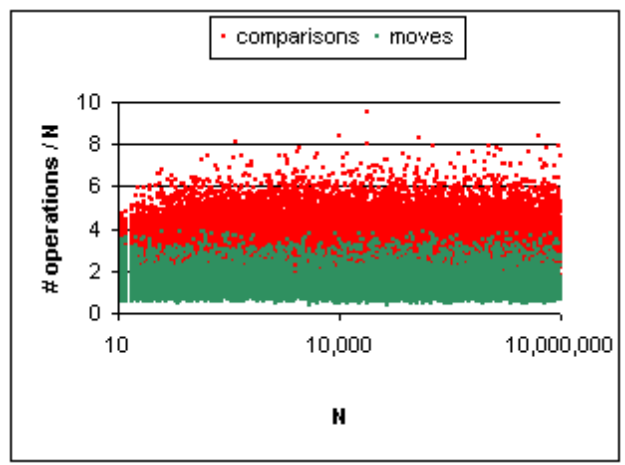

Figure 3-3. Operations by Quickselect

## 4. PARALLEL NATURE

In "The Art of Computer Programming" D.E. Knuth wrote, "Heapsort has the interesting property that its worst case isn't much worse than the average." [8] Although D.E. Knuth's comment was about the heapsort algorithm, Figure 3-1's relatively narrow band of comparison and move operations shows that his observation applies also to the dualheap selection algorithm. This "interesting property" of uniform execution times is even more apparent in the heap construction phases, as shown in Figure 4-1.

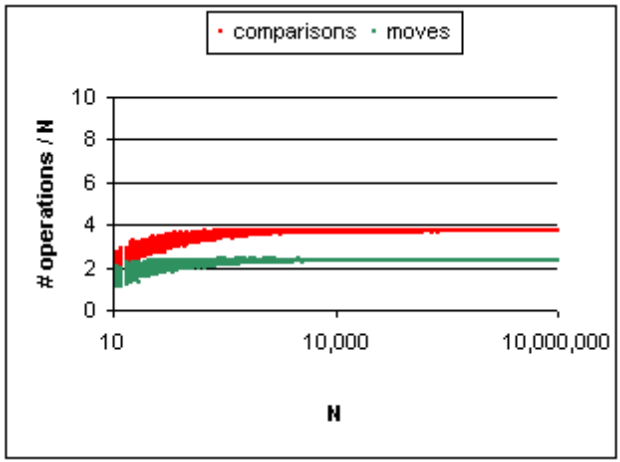

**Figure 4-1. Operations by Heap Construction Phases**

In a coordinated effort in which parallel tasks must wait for the completion of the slowest, this property of uniform execution times boosts overall system performance. And recalling that the partition point of the dualheap selection algorithm is set independently of the values being partitioned, it becomes apparent that all three phases of the dualheap selection algorithm are well-suited for parallel processing: independent processors working on predefined subtasks (independent of the input data), all finishing their tasks at about the same time (rather independent of the input data) so the subtask results can be quickly merged.

## 5. CONCLUSION

Many of our standard algorithm building blocks were initially designed when parallel processing was not much of an issue. Now, the advancement of parallel processing hardware encourages an evolution of some of those designs.

The dualheap selection algorithm offers improvements in efficiency, load balancing and ease of task partitioning. However, the algorithm's worst case analysis remains an open problem. The source code, test cases, spreadsheets and graphs presented in this paper may be a convenient starting point.[10] If you achieve even partial success in analyzing the algorithm, "the author will be pleased to know the details as soon as possible." Ending with that quote from D.E. Knuth [9] is fitting, as the proposed selection algorithm would likely not exist without his encyclopedic coverage of the topic.